\documentclass[showpacs,amsmath,amssymb,aps,prl,floatfix,twocolumn,superscriptaddress,longbibliography]{revtex4-1}
\usepackage{graphicx}
\usepackage{bm}
\usepackage[colorlinks=true,citecolor=blue,linkcolor=red,urlcolor=blue]{hyperref}
\usepackage{dsfont}
\usepackage{color}
\newcommand{\ham}{\mathcal{H}}
\newcommand{\vk}{\mathbf{k}}
\newcommand{\tr}{\mathrm{Tr}}
\newcommand{\setl}[1]{\{\lambda_{#1}\}}

\begin{document}

\title{Locating topological phase transitions using non-equilibrium signatures\\ in local bulk observables}

\author{Sthitadhi Roy}
\affiliation{Max-Planck-Institut f\"ur Physik komplexer Systeme, Dresden 01187, Germany}
\author{Roderich Moessner}
\affiliation{Max-Planck-Institut f\"ur Physik komplexer Systeme, Dresden 01187, Germany}
\author{Arnab Das}
\affiliation{Department of Theoretical Physics, Indian Association for the Cultivation of Science, Kolkata 700032, India}


\begin{abstract}
Topological quantum phases cannot be characterized by local order parameters in the bulk. 
In this work however, we show that non-analytic signatures of a topological quantum critical point do 
remain in local observables in the bulk, and manifest themselves as non-analyticities 
in their expectation values taken over a family of non-equilibrium states generated using a quantum quench protocol. 
The signature can be used for precisely locating the critical points in parameter space. 
A large class of initial states can be chosen for the quench, including finite temperature states. 
We demonstrate these results in tractable models of non-interacting fermions exhibiting topological 
phase transitions in one and two spatial dimensions. We also show that the non-analyticities can be absent 
if the gap-closing is non-topological, i.e., when it corresponds to no phase transition. 
\end{abstract}


\maketitle

\underline{\textit{Introduction }:}
A quantum phase transition is typically  associated with a non-analytic change of the physical properties of the system characterized by a local order parameter measured over the ground state of its Hamiltonian as a function of a tuning parameter.~\cite{Subir_Book,BKC_Book}. 
The signatures of the criticality are also generically not expected to be present in excited eigenstates with finite energy density.
A topological quantum phase transitions (TQPT) by contrast, does not have a \emph{local} order parameter in the bulk which can distinguish the two adjacent phases. 
Different topological phases are characterized by different values of certain topological 
invariants{\cite{HK2010,QZ2011,K2009,RSFL2010,Z1989,TKNN1982} and non-local string 
order parameters \cite{PhysRevLett.59.799,Affleck1988}.

In this work, we present a counterintuitive connection between TQPTs and their signatures in out-of-equilibrium states which are manifestly outside the ground state manifold of the family of Hamiltonians considered.
We show that, {\it local observables in the bulk} 
can show {\it non-analytic} signatures marking the 
ground-state TQPT, where the non-analyticities are observed over a 
family of highly excited states with {\it finite energy densities}. 
Specifically, we work with non-interacting fermionic models where TQPTs can entail conventional QPTs (described by bulk order parameters) 
via transformations (such as Jordan-Wigner) which are crucially \emph{non-local}. Identifying the (transformed) 
bulk order parameter, and hence the transition, turns out to be difficult. Our protocol provides a robust 
prescription for locating the critical point even in such cases via the non-equilibrium footprints described 
in this work. Interestingly, the signature is found to be absent in a case of non-topological gap closing, constructed in this work, which corresponds to no real phase transitions, indicating its ability to distinguish between a true phase transition point and an ``accidental'' gap closing.

Interestingly, manifestation of a QPT via singular signatures far from equilibrium is one most intriguing 
current issue. Our analysis lays down 
a non-equilibrium scheme for the detection of \emph{equilibrium} critical points, and hence is distinct from the 
physics of dynamical quantum phase transitions~\cite{HPK2013,KS2013,VD2015,SSD2015,WSK2016,BH2016,SDPD2016,DSD2016,HuangBalatsky2016}, a priori unrelated to equilibrium critical points~\cite{VD2014}. 
Unlike our non-analytic signatures, these are absent at finite temperatures~\cite{AK2016}. 
Also, our results are distinct from the excited state-QPTs based on DOS effects.~\cite{CCI2008,PCADGR2011,Brandes2013,STP2016}

The concrete models we work with are archetypical models hosting TQPT, namely, the Su-Schrieffer-Heeger 
model\cite{SSH1979} (SSH) and the Kitaev p-wave superconducting chain \cite{K2001} (p-SC) in 1D which 
belong to the symmetry class BDI, and Haldane's honeycomb model\cite{H88} in 2D, 
which belongs to class A. 

Our non-equilibrium protocol~\cite{BDD2015} consists of the following steps. We consider a family of Hamiltonians 
parametrised by a coupling $\lambda$, such that there is a  TQPT as a function of $\lambda$ at the critical point, 
$\lambda = \lambda_{c}$. We start with a state characterised by some initial Hamiltonian $\mathcal{H}(\lambda_i)$ 
(for example, one of its eigenstates or a finite temperature state), and quench it by instantaneously changing the 
parameter from $\lambda_{i}$ to $\lambda_{f}$. Following the quench, the system relaxes to a steady state,  
which can be effectively described by a density matrix, $\rho(\lambda_f)$, diagonal in the eigenbasis of 
$\mathcal{H}(\lambda_f)$ for the purpose of computing expectation-value of local observables on it 
({\it corresponding to the diagonal ensemble})~\cite{RDO2008}. We track the expectation value of local bulk 
observables $\langle {\hat{\cal O}} \rangle = \tr[\hat{\mathcal{O}}\rho(\lambda_{f})]$ as a function of 
$\lambda_{f}.$ We find that $\langle {\hat{\cal O}} \rangle$ reflects the equilibrium topological quantum 
critical points via a non-analyticity in its behavior at $\lambda_f=\lambda_c$. 
A large class of initial states can be used for the quench, since the sufficient condition for obtaining the 
signature turns out to be an occupation gradient across the energy at the gapless modes, which can be achieved 
by controlling the filling fraction or by any finite temperature thermal state. 
This also makes our proposal pertinent for realizations in experiments similar to a recent quench experiment~\cite{flaschner2016observation}.



\underline{\textit{General structure in momentum space}:}
Hamiltonians of the aforementioned one- and two-dimensional systems are translation-invariant and bi-partite in nature,
and can hence be represented in Fourier space by independent two-level systems - each corresponding to a particular momentum mode. 
In terms of the basis vectors $(\vert\vk,A\rangle,\vert\vk,B\rangle)^T$ spanning the Hilbert space of a $k$-mode, the two-level
Hamiltonian is  
\begin{equation}
\ham_\vk(\setl{}) = d_{0,\vk}\mathds{I}_2 + \mathbf{d}_\vk(\setl{})\cdot\bm{\sigma},
\label{eq:ham_tls}
\end{equation}
where $A$ and $B$ denote the two pseudospins (which could be sublattices for bi-partite systems or particle-hole pairs for superconducting systems) 
and the $\sigma$s are the usual Pauli matrices. The Hamiltonian in Eq.\eqref{eq:ham_tls} has two eigenvalues given by 
$\varepsilon_{\pm,\vk}=d_{0,\vk}\pm\vert\mathbf{d}_{\vk}\vert$ and the corresponding eigenvectors are denoted by 
$\vert e_\vk\rangle$ and $\vert g_\vk\rangle$ respectively. We start with a finite temperature mixed density matrix 
corresponding to the initial Hamiltonian $\ham_i = \ham(\setl{i})$ given by $\rho(t=0) = \otimes\prod_\vk \rho_{i,\vk}$, with 
%
$\rho_{i,\vk} = W_{-,\vk}\vert g_{i,\vk}\rangle\langle g_{i,\vk}\vert + W_{+,\vk}\vert e_{i,\vk}\rangle\langle e_{i,\vk}\vert$,
where $W_{\pm,\vk}$ are the Boltzmann weights given by $W_{\pm,\vk} =  e^{-\beta\varepsilon_{\pm,i,\vk}}/(e^{-\beta\varepsilon_{-,i,\vk}}+e^{-\beta\varepsilon_{+,i,\vk}})$. 
Note, that $\mathrm{Tr}[\rho_{i,\vk}]=1$ for every $\vk$ so that the system is half-filled.

Evolution of $\rho_i$ with $\ham_f$, $\rho(t) = e^{-i\ham_f t}\rho_i e^{i \ham_f t}$ after the quench eventually leads to the diagonal ensemble represented by a density matrix 
of the form $\rho_\infty = \otimes\prod_\vk \rho_{\vk,\infty}$, where 
\begin{equation}
\rho_{\infty,\vk} = \frac{1}{2}\left[\mathds{I}_2 + (W_{+,\vk}-W_{-,\vk})\frac{\mathbf{d}_{i,\vk}\cdot\mathbf{d}_{f,\vk}}{d_{i,\vk} d_{f,\vk}^2} \mathbf{d}_{f,\vk}\cdot\bm{\sigma}\right].
\label{eq:rhoinf}
\end{equation}
The expectation value of any local operator $\mathcal{O}$} infinite time can be calculated as 
\begin{equation}
\langle \mathcal{O}\rangle  = \frac{1}{2\pi}\int d\vk~\mathrm{Tr}[\rho_{\infty,\vk}\hat{\cal{O}}_\vk],
\label{eq:obs}
\end{equation}
where the decomposition into the $\vk$-modes is possible because we consider translation invariant operators.

$\langle\mathcal{O}\rangle$ is then studied as a function of $\lambda_f$, the final parameter to which the system is quenched, a non-analyticity is observed when $\lambda_f=\lambda_c$. A natural choice of local observable is the energy difference between the initial and final states, measured with respect to a Hamiltonian corresponding to \emph{any} point $\lambda_m$ in parameter space. Formally, this energy difference is defined as $\Delta E = \tr[\ham_m \rho_\infty] - \tr[\ham_m \rho_0]$. Note that, we keep $\lambda_m$ fixed as we vary $\lambda_f$.  In fact, any local observable which does not commutes with $\mathcal{H}_f$ can capture the non-analyticity, 
The non-analytic signatures persist for any finite temperature initial state (though attenuated as the temperature is increased).
Note that, for conventional QPTs, the energy of ground state of $\mathcal{H(\lambda})$ measured with respect to $\mathcal{H(\lambda})$ itself shows a non-analytic behavior at $\lambda=\lambda_c$


\underline{\textit{TQPT in 1D}:}
This section presents the results for the SSH chain along with a simple physical 
picture to explain the origin of the non-analyticities while the details are relegated to the supplemtary material.

The SSH model is described by the tight-binding Hamiltonian 
%
$\ham_\mathrm{SSH}=-\sum_l [\hat c_{l,A}^\dag \hat c_{l,B} + \lambda\hat c_{l,B}^\dag \hat c_{l+1,A} +\text{h.c.}]$,
%
which corresponds to the reciprocal space Hamiltonian \eqref{eq:ham_tls} with 
%
$d_{\vk}^x = 1+\lambda\cos k;~d_{\vk}^y = \lambda\sin k;~d_\vk^z = 0 = d_{\vk,0}$.
%
The model has two critical points, at $\lambda = \pm1$, with gapless modes at $k=\pi$ and $k=0$ respectively. 
\begin{figure}
\centering
\includegraphics[width=\columnwidth]{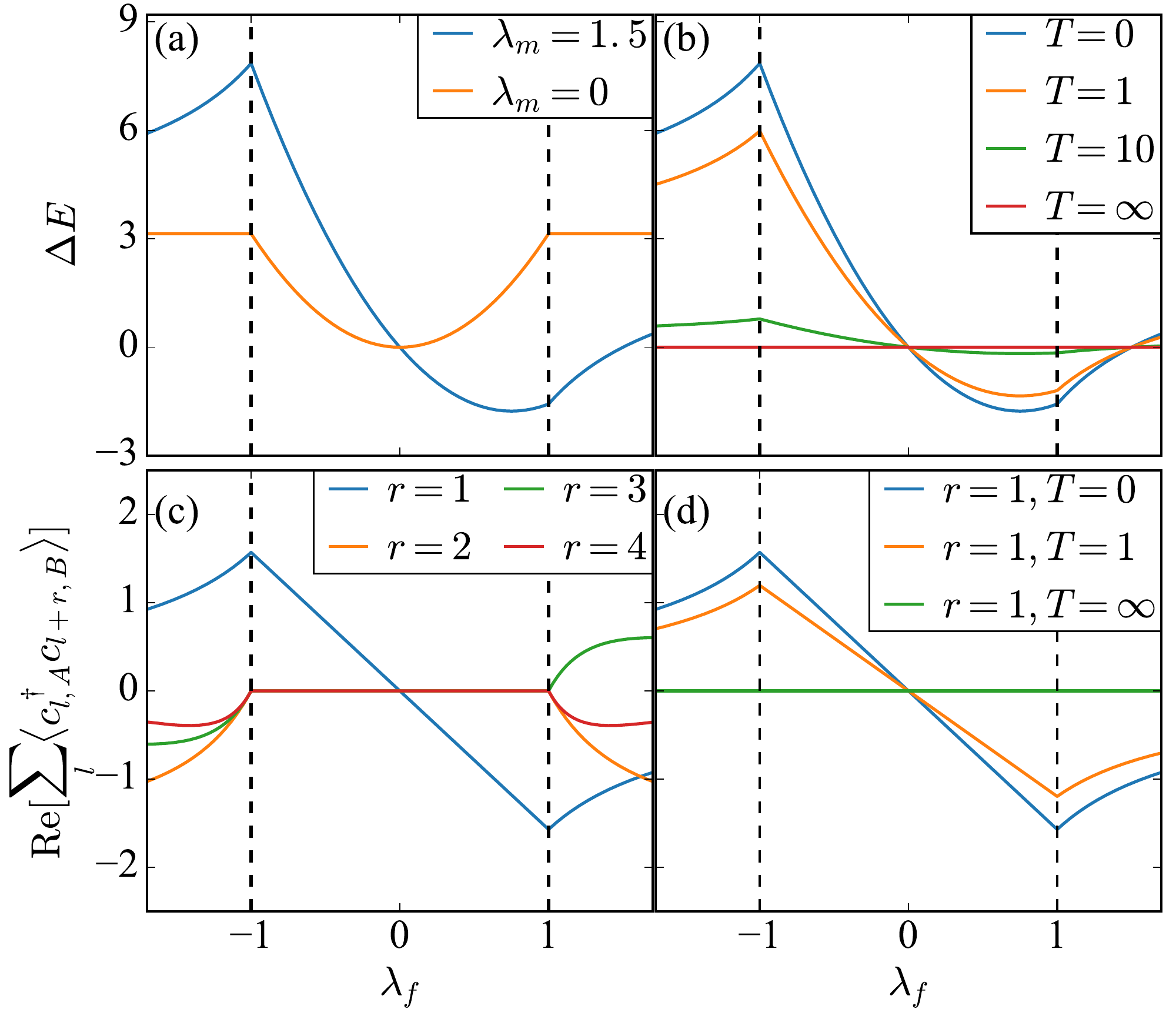}
\caption{ Locating the topological phase transition using {\it non-equilibrium} signatures (the 1D case):
(a) $\Delta E$ for the SSH chain plotted for zero-temperature for $\lambda_i=0$ and two different values of $\lambda_m$. 
The non-analyticities can be clearly seen at the critical points $\lambda_c=\pm1$ denoted by the dashed vertical lines. 
(b) $\Delta E$ for the SSH chain for $\lambda_i=0$ and $\lambda_m=1.5$ for different temperatures. 
The non-analyticities are present at any finite temperature. (c) The off-diagonal (in sublattice space) correlators are 
plotted for the SSH chain which also show non-analyticies at the critical point. (d) The non-analyticities in the local 
correlators also survive the finite temperature ensemble average.}
\label{fig:ssh_signatures}
\end{figure}
The energy difference $\Delta E$ is plotted in Fig.\ref{fig:ssh_signatures}(a)-(b) for different values of the parameters and temperatures, showing the non-analytic behavior at the critical point. 

The non-analyticity in $\Delta E$ as function of $\lambda_f$ appears as a kink at the critical point: the second-derivative of $\Delta E$ with respect to 
$\lambda_f$ diverges at the critical point, as can be  seen by expanding the second-derivative of $\Delta E$ with respect to $\lambda_f$ 
around the gapless mode at the critical point. We take the critical point at $\lambda_f=1$,  and expand in powers of $\kappa = k-\pi$. We find that
\begin{equation}
\left.\frac{\partial^2 (\Delta E(\kappa))}{\partial \lambda_f^2}\right\vert_{\lambda_f=1} = \frac{C_{-2}}{\kappa^2} + C_0 + C_2\kappa^2+\dots.
\label{eq:exp1d}
\end{equation}
 Hence, while the quench protocol populates higher-excited eigenmodes of $\ham_\mathrm{SSH}(\lambda_f)$, the dominant contribution to the non-analyticity of $\Delta E$ comes from the gapless mode $k_c$.

\begin{figure}
\centering
\includegraphics[width=\columnwidth]{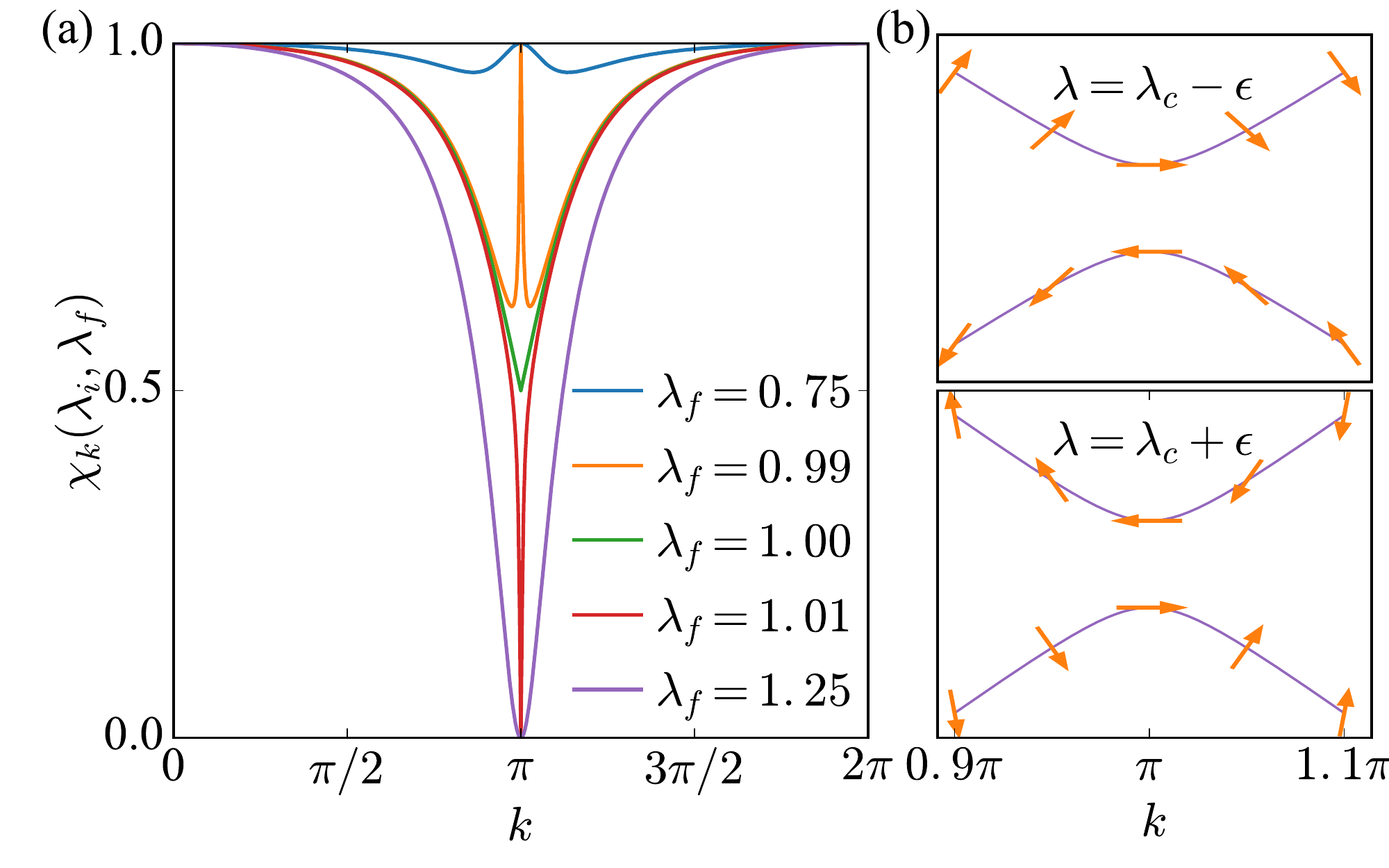}
\caption{(a)The overlap $\chi_k(\lambda_i,\lambda_f)$ plotted as function of $k$ for different values of $\lambda_f$ for a fixed $\lambda_i=0.5$. The overlap at the gapless mode $k=\pi$ stays pinned to one when $\lambda_f$ is on the same side of the critical point as $\lambda_i$, and it switches to zero otherwise. (b) The pseudospin texture is schematically shown on the band dispersion for two values of $\lambda$ on two different sides of the critical point, which shows that the states at the gapless mode become orthogonal across the critical point. For the plots $\epsilon$ is chosen to be 0.1.}
\label{fig:overlap_ssh}
\end{figure}

The mechanism of the non-analyticity can be understood by looking at the mode-by-mode overlap of the initial state with the eigenstates of $\ham_\mathrm{SSH}(\lambda_f)$ across the critical point. For simplicity of illustration, we start with the ground state of $\ham_\mathrm{SSH}(\lambda_i)$ ($\vert\psi(t=0)\rangle=\otimes\prod_k\vert g_k(\lambda_i)\rangle$). We define the overlap as $\chi_k(\lambda_i,\lambda_f) = \vert\langle g_k(\lambda_i)\vert g_k(\lambda_f)\rangle\vert^2$ and plot it as a function of $k$ for different values of $\lambda_f$ in Fig.\ref{fig:overlap_ssh}(a). As long as $\lambda_f$ stays on one side of the critical point, the overlap at the gapless mode ($k_c=\pi$) stays pinned to one, even when it is arbitrarily close to the critical point. However as soon as the critical point is crossed, the overlap jumps to zero discontinuously, where it stays pinned.

This discontinuous jump can be understood via the pseudospin textures (in sublattice space) at the gapless mode across the critical point. Since the SSH chain is a bi-partite system, one can simply compute the pseudospin textures by taking the expectation values of the Pauli matrices with respect to the eigenstates of the Hamiltonian. These textures shown in Fig.\ref{fig:overlap_ssh}(b-c) at the gapless mode ($k=\pi$) for the parameter value $\lambda = 1+\epsilon$ take the form
\begin{equation}
\langle g_\pi(1+\epsilon)\vert\bm{\sigma}\vert g_\pi(1+\epsilon)\rangle=\{\text{sgn}(\epsilon),0,0\}.
\end{equation}
The sign function ensures that across the critical point, the states at the gapless mode are orthogonal to each other, which manifests itself in the overlap switching from one to zero suddenly as the parameter is varied across the critical point.

The above arguments show that the non-analyticity in the observables at the critical point comes from fact that the gapless mode is occupied and the nature of the state at the mode changes in a discontinuous way across the transition. This corroborates the earlier claim that it is not necessary to start from the ground state of $\ham_i$, which also explains why the non-analyticities survive the finite temperature ensemble average. 

In order to show that the non-analyticity hiding in the final density matrix can be captured by almost any local observable we also calculate local correlation functions following the quench. Note that the reciprocal space Hamiltonian of the SSH chain is always restricted to the $x$-$y$ plane in sublattice space and hence any correlator which is diagonal in sublattice space ($\propto\sigma^z$) has zero expectation value. Hence we calculate off-diagonal correlations defined as $\hat{G}_r = \sum_l \langle c_{l,A}^\dagger c_{l+r,B}\rangle$. As expected these correlations also show non-analyticies of the same form as $\Delta E$ and they also survive the finite temperature ensemble averaging as can be seen in Fig.\ref{fig:ssh_signatures}(c)-(d).

To conclude the section we state the results for the case of the Kitaev p-SC chain. Since the Hamiltonian is superconducting, it does not conserve particle number. Hence, apart from the energy and local correlations, the fermion density difference after the quench at $t\rightarrow\infty$ also shows a non-analytic signature of the TQPT\cite{supp}.


\underline{\textit{TQPT in 2D}:}
In this section we investigate the situation in higher spatial dimensions. It is interesting to note that our quench protocol succeeds 
in detecting TQPTs via local bulk observables, whereas it is known that the topological properties of a state does not change following a quantum 
quench although indications of the TQPT can be found by studying the topological edge responses~\cite{AR2015,CCB2015}. We consider Haldane's honeycomb model described by the Hamiltonian 
%
$\mathcal{H}_\mathrm{HM} = -t_1\sum_{\langle i,j\rangle}[c_i^\dagger c_j + \mathrm{h.c}] - t_2\sum_{\langle\langle i,j\rangle\rangle}[e^{i\phi}c_i^\dagger c_j + \mathrm{h.c}]
+ M\sum_i[c_{i,A}^\dagger c_{i,A} - c_{i,B}^\dagger c_{i,B}]$,
where the nearest neighbour hopping is real and the next-nearest neighbour hopping has 
a complex phase $\phi$ encoding the staggered flux through each plaquette. The model 
has critical lines in parameter space given by the relations $M = \pm 3\sqrt{3}t_2 \sin\phi$. 
The reciprocal space Hamiltonian can be written again in terms of the Pauli matrices with coefficients 
%
$d^x_\vk = -t_1(1 + \cos k_2 + \cos(k_2-k_1)),
d^y_\vk = t_1(\sin k_2 + \sin(k_2-k_1)),
d^z_\vk  = M -2t_2\sin\phi(\sin k_2 - \sin k_1 - \sin(k_2-k_1))$,
where $\vk=(k_1,k_2)$ are the reciprocal lattice vectors. The gapless modes corresponding to the two critical lines are at $(\frac{4\pi}{3},\frac{2\pi}{3})$  and $(\frac{2\pi}{3},\frac{4\pi}{3})$.

Apart from the energy difference as before, the other local observable we calculate is the difference between the number of fermions on one sublattice before and after the quench, $\Delta N_A = \tr[\rho_\infty \hat{N}_A] - \tr[\rho_0\hat{N}_A]$, where $\hat{N}_A = \sum_l c_{l,A}^\dagger c_{l,A}$. Note 
that it differs from the staggered occupation operator $\hat{N}_A - \hat{N}_B$ by a constant as the total number of fermions is a constant of motion. In the parameter space of the model, for simplicity, we keep $\phi$ fixed and quench $M$. However, the non-analyticities if present would show up across any quench path across the critical line.
\begin{figure}[htb]
\centering
\includegraphics[width=\columnwidth]{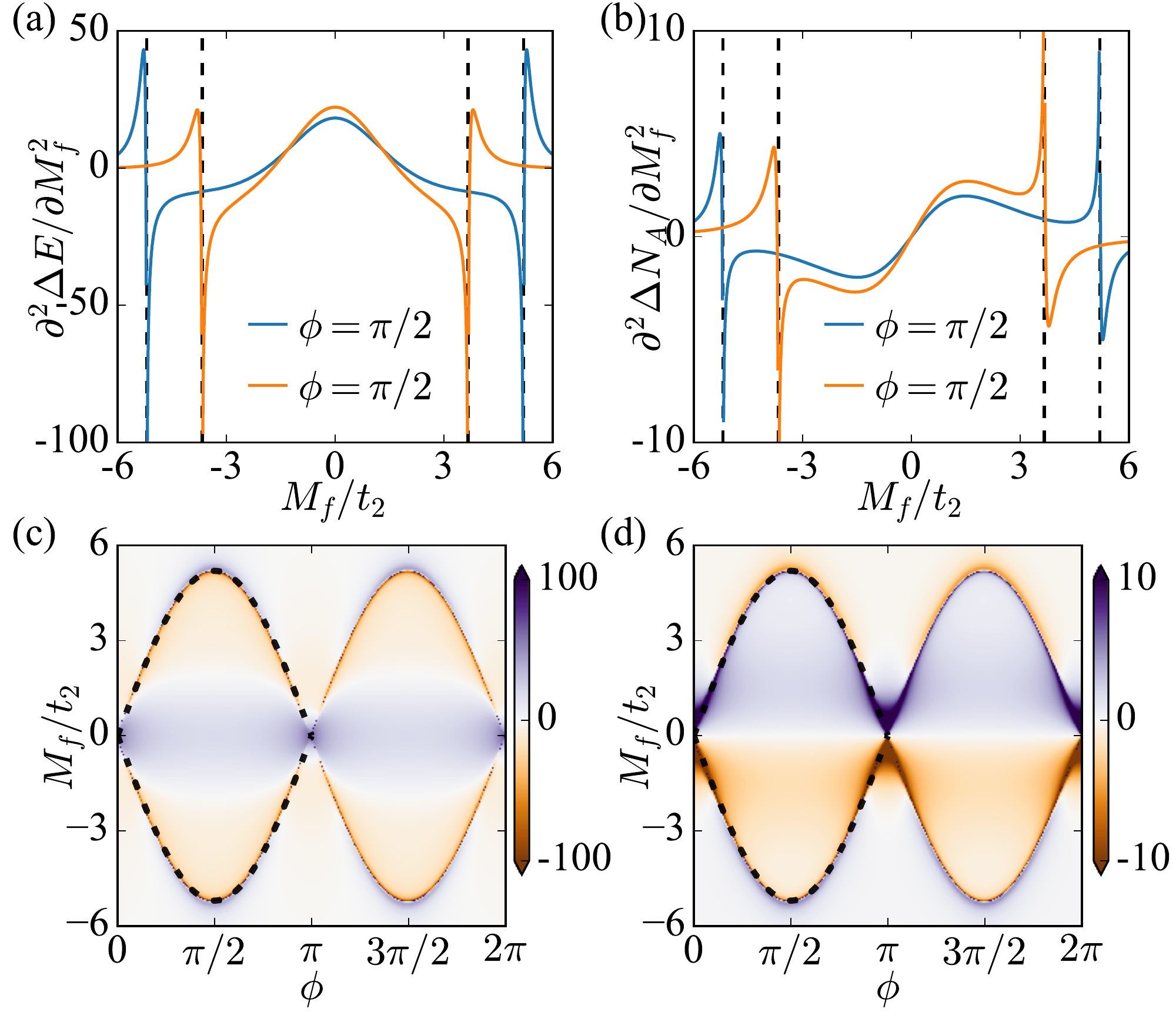}
\caption{Locating the toplogical phase boundary using a non-equilibrium signature in the bulk (the 2D case):
(a)$\partial^2\Delta E/\partial M_f^2$  and (b)$\partial^2\Delta N_A/\partial M_f^2$  diverge at the critcal 
points (marked by the vertical dashed lines) indicating that $\Delta E$ and $\Delta N_A$ have a kink there. 
(c-d) Reconstruction of the phase boundaries using the location of the divergence. The thick dotted lines 
mark the equilibrium transition for the left (symmetric) half of the phase diagram for comparison. }
\label{fig:haldane_signatures}
\end{figure}

The presence or absence of non-analyticities in the expectation values of observables can be studied by looking at the derivatives of these quantities with the final 
value of the quench parameter. As in Eq.\eqref{eq:exp1d} we expand the second derivative of the expectation values of the observables around the gapless mode at the 
critical parameter values. Expressing $k_1 = k_{c,1}+\kappa \cos\theta_\vk$ and $k_2 = k_{c,2}+\kappa \sin\theta_\vk$, we can perform the expansion in powers of 
$\kappa$ 
\begin{equation}
\left.\frac{\partial^2 (\Delta E(\kappa))}{\partial M_f^2}\right\vert_{M_f=3\sqrt{3}t_2\sin\phi} = \frac{C_{-2}}{\kappa^2} +  \frac{C_{-1}}{\kappa} +C_0 + C_2\kappa^2+\dots.
\label{eq:exp2d}
\end{equation}

The nature of the non-analyticity (kink) in the observables depends on the divergence of its second derivative with $\lambda_f$ 
calculated for the gapless mode at $\lambda_f=\lambda_c$, and the integral measure in Eq.\eqref{eq:obs}. It is apparent (see Eqs.\eqref{eq:exp1d} and \eqref{eq:exp2d}) that the non-analyticity is weaker in 2D compared to 1D. 

\begin{figure}[!t]
\centering
\includegraphics[width=\columnwidth]{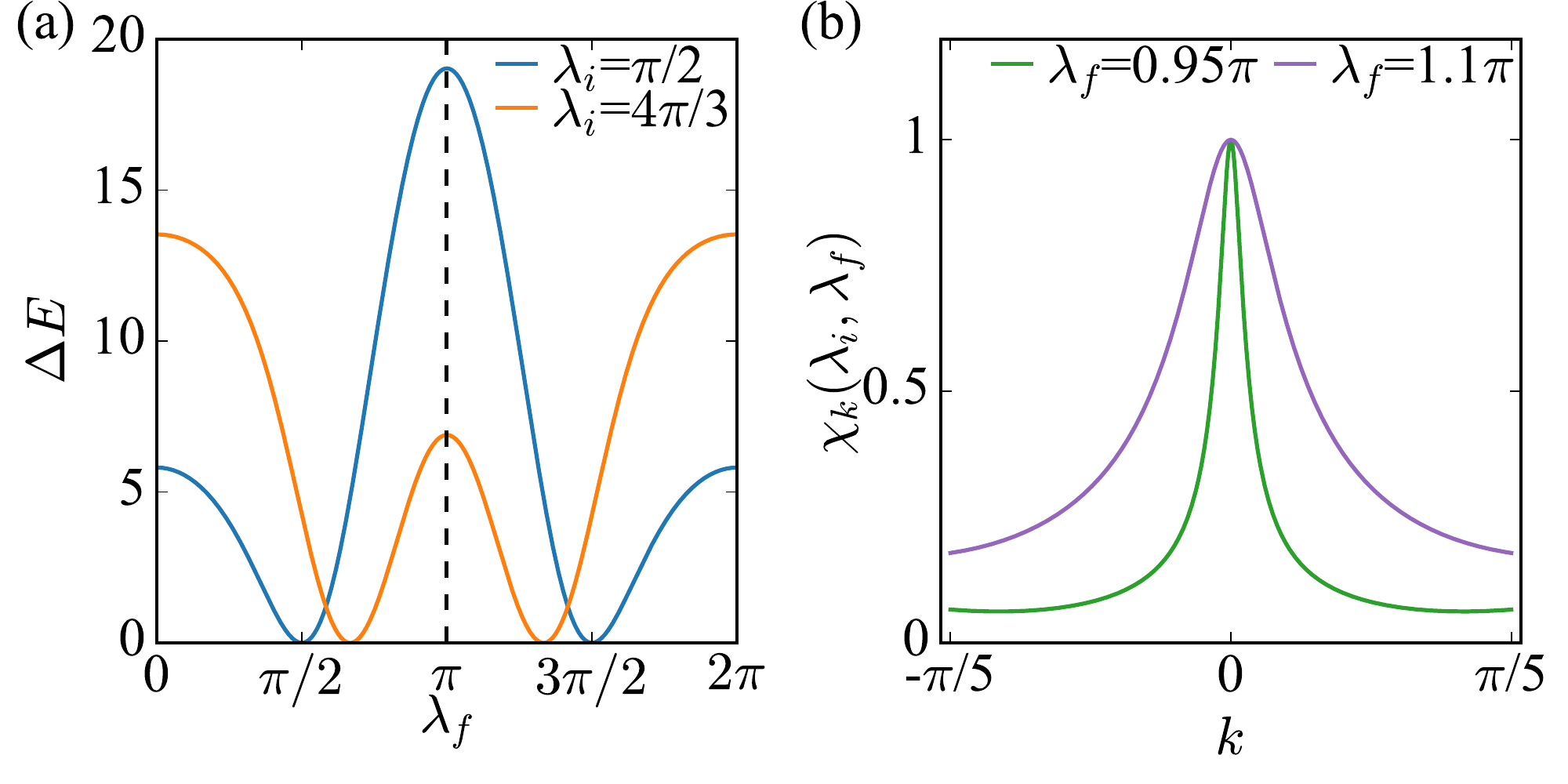}
\caption{(a) The energy absorbed, $\Delta E$, is smooth across the gap-closing for the non-topological case. 
(b) The overlaps at the critical mode stay at one irrespective of which sides of the gap-closing $\lambda_i$ and $\lambda_f$ lie.}
\label{fig:nontop}
\end{figure}

\underline{\textit{Non-topological gap-closings}:} In this section we demonstrate a situation where there exists a 
non-topological linear band touching, which does not give rise to any non-analytic signature. 
Such non-topological gap closings can be studied in a particular two-leg ladder   with complex hoppings. 
Its reciprocal space Hamiltonian takes the form
\begin{equation}
\ham_{NT}(k) = -\sin k\sigma^x + (1-\cos k)\sigma^y + [1+\cos k +2\cos\lambda]\sigma^z.
\label{eq:hamnt}
\end{equation}
The model has a gap closing at $\lambda_c{=}\pi$ at $k_c{=}0$. However, the sign of the effective mass at the gapless mode $2(1+\cos\lambda)$ 
 remains non-negative so that the gap closing does not change the topological properties of the band. Consequently, the pseudospin texture in the 
BZ also does not change suddenly across the gap-closing and in fact at the gapless mode always stays pinned at $\langle\bm\sigma\rangle=\{0,0,1\}$ for 
the lower band. This results in the overlap $\chi_k(\lambda_i,\lambda_f)$ also being pinned to one (see Fig.\ref{fig:nontop}) and hence there is 
no sudden change. This manifests itself in the smooth behavior of $\Delta E$ across the gap closing as shown in Fig.\ref{fig:nontop}. 
Hence, any non-analytic behavior as described in our protocol would signal the presence of a topological quantum critical point.

\underline{\textit{Conclusions and outlook}:}
In this work, we have shown that non-analytic signatures of topological quantum phase transition in non-interacting 
fermionic systems are manifesting in local observables measured on excited states with finite energy densities 
obtained via a quantum quench. 
We have shown that the non-analytic signatures originate from the non-analytic change of the effective 
pseudospin texture at the gapless mode across the critical point, hence the crucial ingredient for observation 
of the signatures is an occupation gradient across the energy of the gapless mode. 
We have found that a gap closing alone is not enough to show the signatures, rather there has to be a phase 
transition (which is a topological one for non-interacting fermions) for the signatures to be present. 

Our findings can be experimentally realised in quantum quench experiments such as in Ref.\cite{flaschner2016observation} designed to study quantum quenches 
in translation invariant two-level systems, just like the ones treated in our work. The two-component spinor corresponding to each $k$-mode forms a unit vector parameterised by two angles on the Bloch sphere, 
which is measured in the experiment as a function of time. Hence, the full information of the quantum state can be extracted and expectation values of any local observable consequently reconstructed.
Although the non-analyticities reported in our work are strictly observed only after infinite evolution times, sufficiently sharp signatures can be expected within the experimentally accessible time-scales of coherent evolution (see supplementary material).

Even though TQPTs in non-interacting fermionic models often correspond to conventional 
phase transitions related via non-local transformations, our protocol does not depend in which representation the transition is topological. 
As these signatures are not present, if the gap-closing in the non-interacting fermionic model does not lead to a change in the topological nature 
of the underlying energy bands participating in the gap-closing, spin-models corresponding to such free fermion models have their 
Hamiltonian parameters restricted in such a way that they are confined to either the ordered or the disordered regimes in their phase diagram. 
Hence, one could also conjecture that any order-disorder phase transition, if it possesses a bonafide single-particle representation, will turn out to be a 
topological one in the non-interacting picture. This is consistent with the general picture proposed in a study for bulk phase transitions ~\cite{BDD2015},
which says that even {\it non-equilibrium} expectation values of local observables in the diagonal ensemble of ${\cal H}(\lambda_{f})$ are smooth functions 
of $\lambda_{f}$ within an {\it equilibrium} phase (defined by local/topological ordering of the ground state), and generally exhibit non-analyticities 
only at the phase boundaries. 
The non-analyticities in the local observables can be traced back to those in the (extensive number of) Lagrange multipliers characterising the generalised Gibbs ensemble describing the steady state of the integrable systems. Whether the few Lagrange multipliers (for instance, the effective temperature) that describe the Gibbs ensemble for interacting systems have a non-analyticity as well, with a concomitant signature in local observables, will require a  detailed investigation which we leave as
an intriguing subject for future work.

\begin{acknowledgments}
{\it Acknowledgments:} SR thanks J-M. St\'ephan and M. Heyl for useful discussions. AD acknowledges support from DST-MPI partner group program ``{\it 
Spin liquids: correlations, dynamics and disorder}" between MPI-PKS (Dresden) and IACS (Kolkata), and the visitor's program of MPI-PKS.
AD also acknowledges S. Bhattacharyya and S. Dasgupta for an earlier collaboration on related topic. 
Authors acknowledge correspondences from Z. Huang pointing out an interesting recent work~\cite{HuangBalatsky2016} on non-equilibrium
signature of topological transitions. This work was in part supported by DFG via SFB 1143.
\end{acknowledgments}


\bibliography{references}


\onecolumngrid

\begin{center}
 \textbf{SUPPLEMENTARY MATERIAL}
\end{center}

\section{Demonstration of the results using the SSH chain}
In this section, we show the workings of the calculations leading to the non-analyticities. For the purpose of demonstration, we choose to show the details of the energy difference calculations and specifically work out the results for the SSH chain.
Following the notation of Eq.~(1) of the main text, the initial and the infinite time density matrices can be expressed as
\begin{eqnarray}
 \rho_{0,\vk} &=& \frac{1}{2}\left[\mathds{I}_2 +(W_{+,\vk}-W_{-,\vk})\frac{\mathbf{d}_{i,\vk}\cdot\bm{\sigma}}{d_{i,\vk}}\right],\\
 \rho_{\infty,\vk} &=& \frac{1}{2}\left[\mathds{I}_2 + (W_{+,\vk}-W_{-,\vk})\frac{\mathbf{d}_{i,\vk}\cdot\mathbf{d}_{f,\vk}}{d_{i,\vk} d_{f,\vk}^2} \mathbf{d}_{f,\vk}\cdot\bm{\sigma}\right].
\end{eqnarray}
Let us choose a local observable for our purpose, which is $\mathcal{H}_m\equiv \mathcal{H}(\lambda_m) =\sum_\vk \mathcal{H}_{m,\vk}=\sum_\vk[d_{0,m,\vk}\mathds{I}_2+\mathbf{d}_{m,\vk}\cdot\bm{\sigma}]$.
Then the energy of the system before and after the quench is given by
\begin{eqnarray}
 E_0 &=& \frac{1}{2\pi}\int d\vk~\mathrm{Tr}[\rho_{0,\vk}\mathcal{H}_{m,\vk}] = \frac{1}{2\pi}\int d\vk~\left[d_{0,m,\vk}+(W_{+,\vk}-W_{-,\vk})\frac{\mathbf{d}_{i,\vk}\cdot \mathbf{d}_{m,\vk}}{d_{i,\vk}}\right], \label{eq:e0}\\
  E_\infty &=& \frac{1}{2\pi}\int d\vk~\mathrm{Tr}[\rho_{\infty,\vk}\mathcal{H}_{m,\vk}] = \frac{1}{2\pi}\int d\vk~\left[d_{0,m,\vk}+(W_{+,\vk}-W_{-,\vk})\frac{\mathbf{d}_{i,\vk}\cdot \mathbf{d}_{f,\vk}}{d_{i,\vk}}\frac{\mathbf{d}_{f,\vk}\cdot \mathbf{d}_{m,\vk}}{d^2_{f,\vk}}\right]\label{eq:einf}.
\end{eqnarray}

From Eqs.~\eqref{eq:e0} and \eqref{eq:einf}, it is easy to see that $\Delta E = E_\infty-E_0$ is zero for $\lambda_f=\lambda_m$, one of the results stated in the main text.

For the SSH chain, $\Delta E$ turns out to be 
\begin{equation}
\Delta E = \int_0^{2\pi} dk\frac{(W_{+,\vk}-W_{-,\vk})({\lambda_f}- {\lambda_i}) ( {\lambda_f}- {\lambda_m}) \sin ^2 k }{\left( {\lambda_f}^2+2  {\lambda_f} \cos  k +1\right) \sqrt{ {\lambda_i}^2+2  {\lambda_i} \cos  k +1}}.
\label{eq:eabsssh}
\end{equation}

For simplicity of presentation, we take $\lambda_i=0=\lambda_m$ (orange curve in Fig. 1(a) of main text). With these parameters, Eq.~\eqref{eq:eabsssh} turns out to be of the form
\begin{equation}
 \Delta E = \frac{\pi}{2}\left[(\lambda_f^2+1) - (\lambda_f+1)^2\left\vert\frac{\lambda_f-1}{\lambda_f+1}\right\vert\right]=
 \begin{cases}
  \pi\lambda_f^2; ~~\vert\lambda_f\vert<1\\

  \pi; ~~~~~\vert\lambda_f\vert>1\\\end{cases},
\end{equation}
which clearly shows that the first derivative of $\Delta E$ with respect to $\lambda_f$ is discontinuous at the critical points $\lambda_c=\pm 1$, and hence the non-analyticity.

For other values of $\lambda_i$ and $\lambda_m$, a closed form expression is difficult to write, however a pole-structure analysis of the integral in Eq.~\ref{eq:eabsssh}, similar to Ref.~[28] of the main text shows the mathematical origin of the non-analyticities.

\section{Signatures at finite times}
\begin{figure}[htb]
\centering
\includegraphics[width=0.75\columnwidth]{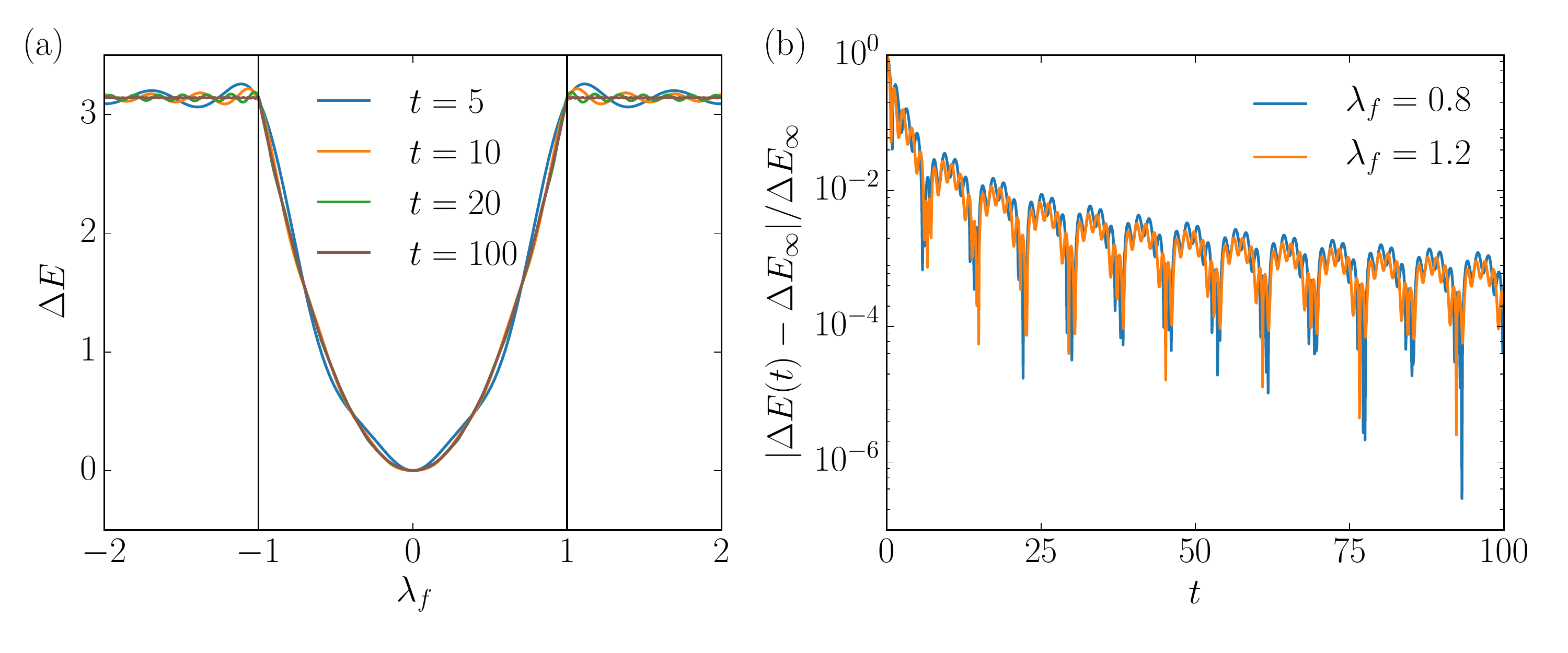}
\caption{ ~(a) $\Delta E$ as a function of $\lambda_f$  for the SSH chain at finite times for $\lambda_i=0$. The times are quoted in units of inverse hopping. The result for $t=100$ (dashed) already shows a rather sharp signature of the critical point similar to the Fig.~1(a) of the main text. At finite times, the critical point can still be clearly located by the qualitative change of the $\lambda_f$ dependence of $\Delta E$. (b) Approach to the diagonal ensemble value. The relative difference between the finite time and the diagonal ensemble results is plotted against time.}
\label{fig:finitet}
\end{figure}

As discussed in the main text, our results can be experimentally verified via quantum quench experiments similar to the one in Ref.~[31]. Our results show that the non-analyticities at the critical points appear in the limit of inifite time following the quench as the density matrix describing the state appoaches the one corresponding to the diagonal ensemble. At finite times, the non-analyticities are dressed by the off-diagonal contributions. However, explicit calculations show that even at finite times accessible in experiments, the critical points can be identified even though the cusp present at infinite times is rounded off. For instance, in Fig.~\ref{fig:finitet} we show $\Delta E$ for the SSH chain at different times $t$ (measured in units of inverse hopping). At time scales achieveable in the experiment in Ref.[31], namely $t=10$ and $t=20$, one can identify the critical points from the change in nature of the functions around $\lambda_f=\pm1$. The emerging cusp is already clearly discernible at $t=100$.

\section{Results for Kitaev p-wave chain}
\begin{figure}[htb]
\centering
\includegraphics[width=0.5\columnwidth]{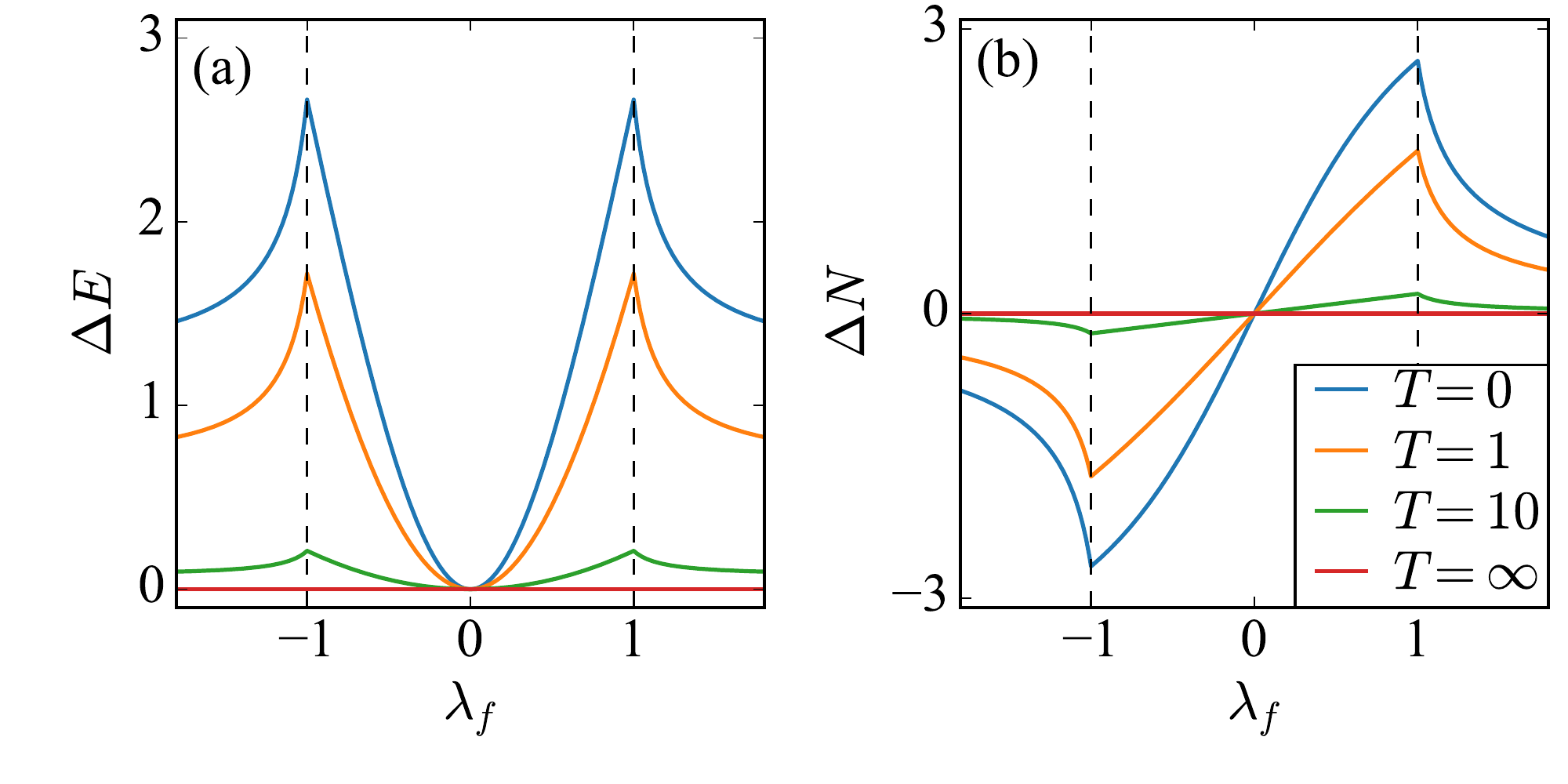}
\caption{ ~(a) $\Delta E$ calculated for the Kitaev p-SC chain shows the non-analyticities at the critical points (denoted by the black dashed lines) at different temperatures. 
For simplicity we use $\lambda_m = \lambda_i=0$. (b) The fermion number difference for the same model calculated using Eq.\eqref{eq:deltanpsc} also shows non-analyticities at the critical 
points.}
\label{fig:psc_signatures}
\end{figure}
In this section we show that the non-analytic signatures of TQPT are present in the Kitaev p-SC chain. The Hamiltonian of the model is given by 
\begin{equation}
\begin{split}
\ham_\mathrm{p-SC} = -\sum_i [c_i^\dagger c_{i+1}& + \mathrm{h.c}] + \lambda\sum_i  c_i^\dagger c_i +\\
& \sum_i [\Delta_\mathrm{SC} c_i^\dagger c_{i+1}^\dagger + \mathrm{h.c}],
\end{split}
\label{eq:hampw}
\end{equation}
where $\lambda$, the chemical potential is our quench parameter and $\Delta_\mathrm{SC}$ is the superconducting order parameter. The coefficients of the Pauli matrices in reciprocal space Hamiltonian are 
%
%
%
\begin{equation}
d_{\vk}^x = 0 = d_{\vk,0};~d_{\vk}^y = \Delta_\mathrm{SC}\sin k;~d_\vk^z  = -\cos k-\lambda,
\label{eq:hamkpsc}
\end{equation}
where the basis now is $(\vert k \rangle,\vert -k\rangle)^T$. This model has a phase transition between a topological superconducting phase and a normal superconducting phase at $\lambda_c=\pm$. The model being a superconducting one does not conserve fermion number which naturally suggests a local observable which is experimentally relevant, namely the difference in the number of fermions before and after the quench, $\Delta N = \tr[\hat{N}\rho_\infty] - \tr[\hat{N}\rho_0]$, where $\hat{N}$ is the total fermion number operator. $\Delta N$ turns out to be
\begin{equation}
\Delta N  = \int_0^{2\pi} dk ~ (W_{+,\vk}-W_{-,\vk})\frac{\mathbf{d}_{i,\vk}\cdot\mathbf{d}_{f,\vk}} {d_{i,\vk} d_{f,\vk}^2}d_{f,\vk}^z,
\label{eq:deltanpsc}
\end{equation}
where the vectors $\mathbf{d}_\vk(\lambda)$ are given by Eq.\eqref{eq:hamkpsc}.
The difference in fermion number also shows a sharp kink at the critical points like the energy difference. As before these non-analyticities are robust towards a finite temperature ensemble average. These non-analyticities can be seen in Fig.\ref{fig:psc_signatures}.

\end{document}